\documentclass[aps,prl,twocolumn,showpacs]{revtex4}
\usepackage{graphicx}

\begin{document}

\title{On the rationality of the OPERA experiment as a signal of Lorentz violation}

\newcommand*{\PKU}{School of Physics and State Key Laboratory of Nuclear Physics and
Technology, \\Peking University, Beijing 100871,
China}\affiliation{\PKU}
\newcommand*{\CHEP}{Center for High Energy
Physics, Peking University, Beijing 100871,
China}\affiliation{\CHEP}

\author{\surname{Zhou} Lingli}
\email[email:]{ zhoull@pku.edu.cn}\affiliation{\PKU}
\author{Bo-Qiang Ma}
\email[email:]{
mabq@pku.edu.cn}\affiliation{\PKU}\affiliation{\CHEP}

%\affiliation{School of Physics and State Key Laboratory of Nuclear
%Physics and Technology, Peking University, Beijing 100871, China}

\date{\today}

\begin{abstract}
We show that the superluminal muon neutrinos in the recent OPERA
experiment can exist theoretically. The refutation of the OPERA
experiment from some theoretical arguments is not universally valid,
but resulting from some implicit assumptions. Our argument can
accommodate both the OPERA experiment for superluminal neutrinos and
the ICARUS experiment of no evidence for the analogues Cherenkov
radiation of muon neutrinos from CERN to the LNGS.
\end{abstract}

\keywords{neutrino speed, light speed, Lorentz violation}
\pacs{11.30.Cp, 12.60.-i, 14.60.Lm, 14.60.St}%LV\SMS\14.60.Lm Ordinary neutrinos\14.60.St Non-standard-model neutrinos

\maketitle

Recently, the OPERA collaboration reported that the speed of muon
neutrinos is larger than the vacuum light speed by a factor of
$10^{-5}$~\cite{opera11}. Such observation has been
suggested~\cite{MaZhouOpera11,QinMaOperaSME11} as a signal for
Lorentz violation within several existing
frameworks~\cite{SMSmpla10,SMScpc11,SME98} of Lorentz violation, and
the corresponding Lorentz violation parameters of muon neutrinos are
estimated to be of the order of
$10^{-5}$~\cite{MaZhouOpera11,QinMaOperaSME11}. Cohen and Glashow
argued that if the Lorentz violation of the OPERA experiment is of
$10^{-5}$, the high energy muon neutrinos exceeding tens of GeVs can
not be detected by the Gran Sasso detector, mainly because of the
energy-losing process $\nu_\mu\rightarrow\nu_\mu+e^{+}+e^{-}$
analogous to Cherenkov radiations through the long baseline about
730~km~\cite{Glashow11}. Bi {\it et al.} also argued that the
Lorentz violation of muon neutrinos of order $10^{-5}$ will forbid
kinematically the production process of muon neutrinos
$\pi\rightarrow \mu + \nu_\mu$ for muon neutrinos with energy larger
than about 5~GeV~\cite{Bi11}. Such arguments put up a strong
challenge to the rationality of the OPERA experiment and the
consequent suggestion to attribute the OPERA experiment as a signal
of Lorentz violation. In the following, with a Lorentz violation
framework from basic considerations~\cite{SMSmpla10,SMScpc11}, we
discuss the three Cohen-Glashow processes first, and consider also
the production processes of Bi {\it et al.}.

A response to Cohen, Glashow and Bi {\it et al.} is offered in
Ref.~\cite{Smolin11}, where Amelino-Camelia, Freidel,
Kowalski-Glikman and Smolin argued that the energy threshold for the
anomalously Cherenkov analogous process
$\nu_\mu\rightarrow\nu_\mu+e^{+}+e^{-}$ makes physics
observer-dependent. They pointed out that the deformed Lorentz
transformation can avoid the problems brought about by these
arguments. In this paper we propose an alternative argument for the
rationality of the OPERA result as a signal for Lorentz violation,
and we do not use deformed Lorentz transformations. From the point
of view of the Lorentz violation framework, the arguments of
Ref.~\cite{Glashow11} are just one of theoretical possibilities from
some implicit assumptions, which are not easily seen obviously, such
as that neutrinos have the fixed Lorentz violation when the three
Cherenkov-like processes happen (this is the most implicit there),
that all the three generations of neutrinos have the same Lorentz
violation and that the Lorentz violation is described enough by a
scalar parameter. When we work in any Lorentz invariance violation
frameworks, such as the standard model supplement
(SMS)~\cite{SMSmpla10,SMScpc11} and the standard model extension
(SME)~\cite{SME98}, the previously well known physical regulations
gotten in the case of Lorentz invariance become subtle. Since the
Lagrangian in the Lorentz violation framework does not contain
explicitly space-time coordinates, so the translation invariance of
space-time of the Lagrangian is kept. Then the 4-momenta
conservation law is also applicable. The question whether the
process is kinematically forbidden can be discussed. We assume that
the 4-momenta $p$ of real particles are time-like, i.e. $p^2\geq0$
and the time-component of $p$ is non-negative.

The kinematic limit for the  process
$\nu_\mu\rightarrow\nu_\mu+e^{+}+e^{-}$ is
\begin{equation}\label{kin_limit_glashow}
|p_{\nu_\mu,i}|\geq|p_{\nu_\mu,f}|+|p_{e^{+}}|+|p_{e^{-}}|.
\end{equation}
Without Lorentz violation, the mass-energy relation is  $p^2=m^2$.
Then Eq.~(\ref{kin_limit_glashow}) becomes
\begin{equation}\label{kin_limit_glashow_noLV}
m_{\nu_\mu}\geq m_{\nu_\mu}+2m_{e}.
\end{equation}
It is obvious that Eq.~(\ref{kin_limit_glashow_noLV}) can not be
satisfied. So the Cherenkov analogous process
$\nu_\mu\rightarrow\nu_\mu+e^{+}+e^{-}$ is forbidden kinematically
in space-time in the case of no Lorentz violation. We use here a
Lorentz violation framework, e.g. the SMS~\cite{SMSmpla10,SMScpc11},
in which there is a replacement of the ordinary partial
$\partial_{\alpha}$ and the covariant derivative $D_{\alpha}$ by
$\partial^{\alpha} \rightarrow M^{\alpha\beta}\partial_{\beta}$ and
$D^{\alpha}\rightarrow M^{\alpha\beta}D_{\beta}$, where
$M^{\alpha\beta}$ is a local matrix. We separate it to two matrices
like $M^{\alpha \beta}=g^{\alpha \beta}+\Delta^{\alpha \beta}$,
where $g^{\alpha\beta}$ is the metric tensor of space-time. Since
$g^{\alpha\beta}$ is Lorentz invariant, $\Delta^{\alpha \beta}$
contains all the Lorentz violating degrees of freedom from
$M^{\alpha \beta}$. Then $\Delta^{\alpha \beta}$ brings new terms
violating Lorentz invariance in the standard model and is called
Lorentz invariance violation matrix therefore. The mass-energy
relation for electrons and neutrinos
becomes~\cite{MaZhouOpera11,SMSmpla10}
\begin{equation}\label{disp_relation}
p^2+g_{\alpha\mu}\Delta^{\alpha\beta}\Delta^{\mu\nu}p_{\beta}p_{\nu}
+2\Delta^{\alpha\beta}p_{\alpha}p_{\beta}-m^2=0.
\end{equation}
So
\begin{equation}\label{mass_energy_LV_fermion}
p^2=m^2+\lambda(\Delta,p),
\end{equation}
where
\begin{eqnarray}
\lambda(\Delta,p)&&\equiv
-g_{\alpha\mu}\Delta^{\alpha\beta}\Delta^{\mu\nu}p_{\beta}p_{\nu}
-2\Delta^{\alpha\beta}p_{\alpha}p_{\beta}\nonumber\\
&&=-p^tG\Delta^t G \Delta G p-2p^tG\Delta G p.
\end{eqnarray}
At the last step of the above formulas, matrix notation is taken for
simplicity. Here, $p^\alpha\equiv p$,
$g_{\alpha\beta}=\textrm{diag}(1,-1,-1,-1)\equiv G$,
$\Delta^{\alpha\beta}\equiv\Delta$. When the Lorentz violation
matrix $\Delta^{\alpha\beta}$ is diagonal, e.g.
$\Delta^{\alpha\beta}=\textrm{diag}(\eta,\xi,\xi,\xi)$,
Eq.~(\ref{disp_relation}) becomes
$E^2=(|\vec{p}|^2(1-2\xi+\xi^2)+m^2)/(1+2\eta+\eta^2)$. We write it
as the conventional form $E^2=|\vec{p}|^2 c_A^2+(m^{'})^2c^4_A$,
where $c_A^2\equiv(1-2\xi+\xi^2)/(1+2\eta+\eta^2)$ and $c_A$ is
denoted as the maximal attainable velocity of type $A$ particles in
Refs.~\cite{Coleman99,Glashow11}. So the effects of the maximal
attainable velocity of particles can also be provided by a diagonal
Lorentz violation matrix $\Delta^{\alpha\beta}$ of particles here.
As Cohen and Glashow did, we neglect the Lorentz violation of
electrons here. Now, Eq.~(\ref{kin_limit_glashow}) reads
\begin{equation}\label{kin_limit_glashow_LV}
\sqrt{m^2_{\nu_\mu}+\lambda(\Delta_{\nu_\mu,i},p_{\nu_\mu,i})} \geq
\sqrt{m^2_{\nu_\mu}+\lambda(\Delta_{\nu_\mu,f},p_{\nu_\mu,f})}
+2m_{e}.
\end{equation}
The difference between 4-momenta $p_{\nu_\mu,i}$ and $p_{\nu_\mu,f}$
can be notated as a Lorentz transformation $R$:
$p_{\nu_\mu,f}=Rp_{\nu_\mu,i}$. When the Lorentz violation matrix
$\Delta_{\nu_\mu,i}$ of the initial muon neutrino and
$\Delta_{\nu_\mu,f}$ of the final/outgoing muon neutrino are same
and both are diagonal, e.g.
$\Delta_{\nu_\mu}=\textrm{diag}(\eta,\xi,\xi,\xi)$ and it is SO(3)
invariant, a threshold energy for muon neutrinos can be gotten.
Neglecting the mass $m_{\nu_\mu}$ of muon neutrinos and taking the
diagonal form for both $\Delta_{\nu_\mu,i}$ and $\Delta_{\nu_\mu,f}$
in Eq.~(\ref{kin_limit_glashow_LV}), we can get the same threshold
energy for initial muon neutrinos as that in Ref.~\cite{Glashow11}:
$E_\textrm{th}=2m_e/\sqrt{\delta}$, where $\delta\equiv
-2(\eta+\xi)$. If the energy of muon neutrinos is higher than the
threshold energy, the process becomes allowed kinematically and the
process happens. Now, it is clear that the process is kinematically
permitted and a threshold energy of Ref.~\cite{Glashow11} for
initial muon neutrinos can be gotten are based on conditions that
$\Delta_{\nu_\mu,i}=\Delta_{\nu_\mu,f}$ and that
$\Delta_{\nu_\mu,i}, \Delta_{\nu_\mu,f}$ are diagonal, i.e. the
Lorentz violation of muon neutrinos is assumed to have a fixed form.
%without frame dependence.

\begin{figure}
  \includegraphics[width=.45\textwidth]{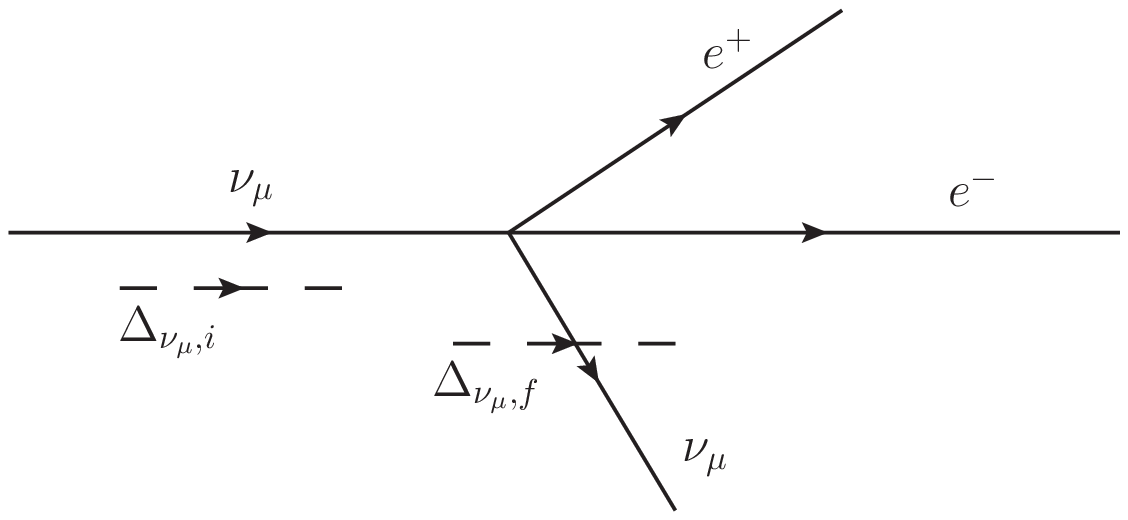} \includegraphics[width=.45\textwidth]{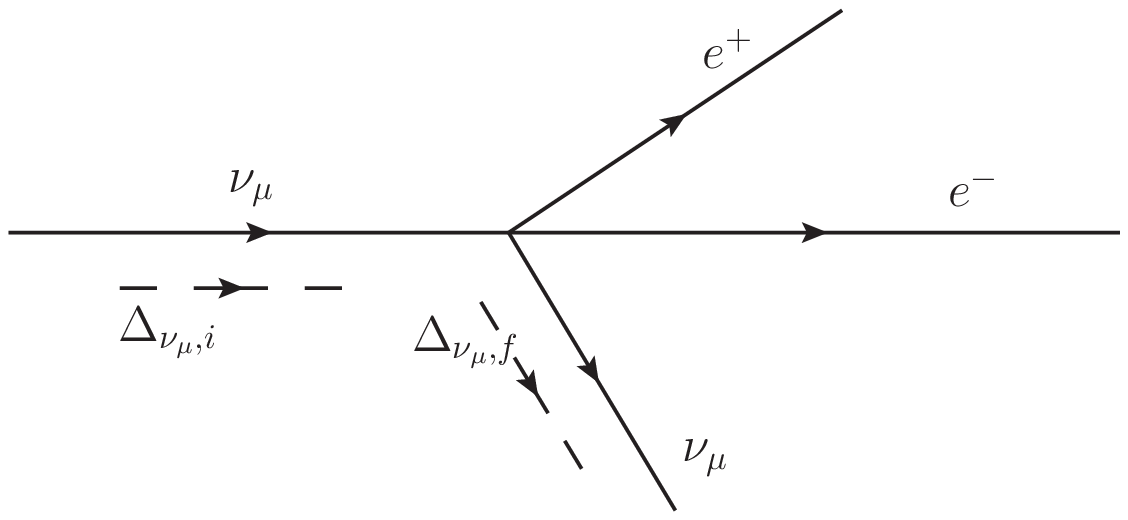}\\
  \caption{The fixed form of Lorentz violation for muon neutrinos is illustrated on the upper-panel.
  The covariant Lorentz violation for muon neutrinos is illustrated on the lower-panel.}\label{fixed_covariant_LV}
\end{figure}

What will happen if the Lorentz violation matrix $\Delta_{\nu_\mu}$
of muon neutrinos is covariant with the corresponding momentum
$p_{\nu_\mu}$: $\Delta_{\nu_\mu,f}=R\Delta_{\nu_\mu,i}R^t$? We find
that
\begin{eqnarray}
&&\lambda(\Delta_{\nu_\mu,f},p_{\nu_\mu,f})=\lambda(R\Delta_{\nu_\mu,i}R^t,R p_{\nu_\mu,i})\nonumber\\
&&=-p^t_{\nu_\mu,i} R^tG R\Delta^t_{\nu_\mu,i} R^t G  R \Delta_{\nu_\mu,i} R^t G R p_{\nu_\mu,i}\nonumber\\
&&-2p^t_{\nu_\mu,i} R^t G R \Delta_{\nu_\mu,i} R^t G R p_{\nu_\mu,i}\nonumber\\
&&=-p^t_{\nu_\mu,i} G \Delta^t_{\nu_\mu,i}G\Delta_{\nu_\mu,i} G
p_{\nu_\mu,i}
-2p^t_{\nu_\mu,i} G \Delta_{\nu_\mu,i} G p_{\nu_\mu,i}\nonumber\\
&&=\lambda(\Delta_{\nu_\mu,i},p_{\nu_\mu,i}),\nonumber
\end{eqnarray}
where $R^tGR=G$ is used. Now we get
$\lambda(\Delta_{\nu_\mu,f},p_{\nu_\mu,f})=\lambda(\Delta_{\nu_\mu,i},p_{\nu_\mu,i})$.
Eq.~(\ref{kin_limit_glashow_LV}) becomes $0\geq2m_e$, and it can not
be satisfied therefore. Since Eq.~(\ref{kin_limit_glashow_LV}) is
equivalent to $0\geq2m_e$, Eq.~(\ref{kin_limit_glashow_LV}) is
equivalent to Eq.~(\ref{kin_limit_glashow_noLV}). So the process
$\nu_\mu\rightarrow\nu_\mu+e^{+}+e^{-}$ is still forbidden
kinematically. With the same discussion, the processes
$\nu_\mu\rightarrow\nu_\mu+\gamma$ and
$\nu_\mu\rightarrow\nu_\mu+\nu_e+\bar{\nu}_e$ are kinematically
forbidden too. The corresponding physical pictures of the fixed
Lorentz violation and the covariant Lorentz violation are shown in
Fig.~\ref{fixed_covariant_LV}. Since Eq.~({\ref{disp_relation}}) is
Lorentz covariant, it is more natural that the Lorentz violation
matrix $\Delta_{\nu_\mu}$ of muon neutrinos is covariant. On the
other hand, for the Cherenkov case, the electromagnetic radiations
result from the polarizing of the corresponding medium by the
superluminal charged particles, and the response of the medium to
the superluminal particles is covariant with the momentum of these
particles. It is more appropriate that the Lorentz violation of muon
neutrinos is emergent and covariant with muon neutrinos for the
process $\nu_\mu\rightarrow\nu_\mu+e^{+}+e^{-}$ in space-time.

\begin{figure}
  \centering
  \includegraphics[width=.4\textwidth]{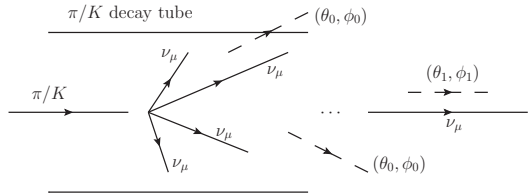}\\
  \caption{A brief illustration of the OPERA muon neutrino beam.}\label{opera_illustration}
\end{figure}

Now we consider the production process of muon neutrinos:
$X\rightarrow\mu +\nu_\mu$, $X=\pi,K$. The corresponding kinematic
limit condition is
\begin{equation}\label{kin_limit_bi}
|p_{X}|\geq|p_{\mu}|+|p_{\nu_\mu}|.
\end{equation}
Without Lorentz violation of muon neutrinos,
Eq.~(\ref{kin_limit_bi}) is equivalent to $m_X\geq
m_\mu+m_{\nu_\mu}$, which can be satisfied. So the process happens
in the case of no Lorentz violation. When the Lorentz violation of
muon neutrinos is considered, from
Eq.~(\ref{mass_energy_LV_fermion}), Eq.~(\ref{kin_limit_bi}) reads
\begin{equation}\label{kin_limit_bi_LV}
m_X\geq
m_\mu+\sqrt{m^2_{\nu_\mu}+\lambda(\Delta_{\nu_\mu},p_{\nu_\mu})},
\end{equation}
where just the Lorentz violation of neutrinos is considered. Even if
the Lorentz violation of muon neutrinos in the OPERA experiment is
of $10^{-5}$, Eq.~(\ref{kin_limit_bi_LV}) can still be satisfied,
i.e. the production process is still kinematically allowed. The
arguments of Ref.~\cite{Bi11} that the Lorentz violation of the
OPERA forbids the production process mainly results from the fact
that a scalar Lorentz violation parameter there can not provide more
details about Lorentz violation in space-time. The Lorenz violation
is related to all dimensions of space-time. The Lorentz violation
matrix $\Delta_{\nu_\mu}$ of muon neutrinos has tens of degrees of
freedom. Calculating $\lambda(\Delta_{\nu_\mu},p_{\nu_\mu})$, we
notate that
$p_{\nu_\mu}^\alpha=(E,|\vec{p}|\Omega^1,|\vec{p}|\Omega^2,|\vec{p}|\Omega^3)$,
where
$\Omega^i=(\sin{\theta}\cos{\phi},\sin{\theta}\sin{\phi},\cos{\theta})$.
We can rewrite $\lambda(\Delta_{\nu_\mu},p_{\nu_\mu})$ for
simplicity as
\begin{equation}
\lambda(\Delta_{\nu_\mu},p_{\nu_\mu})=\alpha E^2+ \beta E|\vec{p}|+
\gamma |\vec{p}|^2,
\end{equation}
where $\alpha=-2\Delta^{00}+O(\Delta)$,
$\beta=4\Delta^{(0i)}\Omega^i+O(\Delta)$, and
$\gamma=-2\Delta^{ij}\Omega^i\Omega^j+O(\Delta)$. So $\alpha$ is
angle independent, and $\beta$ and $\gamma$ are angle dependent. The
same indices $i,j$ mean a summation as usual. Around the energy
range of OPERA of $\sim$17~GeV, $m_{\nu_\mu}\ll E$. So the terms of
$m_{\nu_\mu}$ are neglected. From
Eq.~(\ref{mass_energy_LV_fermion}), we get that $E=E(|\vec{p}|)$.
Then
\begin{displaymath}
\lambda(\Delta_{\nu_\mu},p_{\nu_\mu})=(\alpha+\beta+\gamma) E^2,
\end{displaymath}
which is approximated to the first order of $\alpha$, $\beta$ and
$\gamma$. There exist directions $(\theta_0,\phi_0)$ such that
$\alpha+\beta(\theta_0,\phi_0)+\gamma(\theta_0,\phi_0)\simeq0$. So
$\lambda(\Delta_{\nu_\mu},p_{\nu_\mu})\big|_{\theta_0,\phi_0}\simeq
0$~MeV$^2$, and Eq.~(\ref{kin_limit_bi_LV}) is satisfied, i.e. the
production processes $\pi/K\rightarrow\mu +\nu_\mu$ are
kinematically allowed in the case of Lorentz violation of $\sim
10^{-5}$ for the OPERA muon neutrinos. From
Eq.~(\ref{mass_energy_LV_fermion}), we can also get
\begin{equation}\label{speed_anomaly}
v_{\nu_\mu}\equiv
\frac{dE}{d|\vec{p}|}=1+\frac{1}{2}\left(\alpha+\beta+\gamma\right)+f(\alpha,\beta,\gamma),
\end{equation}
which is approximated to the second order. And
$f(\alpha,\beta,\gamma)=(3\alpha^2+\beta^2-\gamma^2+4\alpha\beta+2\alpha\gamma)/8$.
Muon neutrinos propagate to Gran Sasso from CERN in direction
$(\theta_1,\phi_1)$. Generally,
$(\theta_1,\phi_1)\neq(\theta_0,\phi_0)$. Then
Eq.~(\ref{speed_anomaly}) becomes
\begin{eqnarray}
&&v_{\nu_\mu}\big|_{\theta_1,\phi_1}=1+\frac{1}{2}\left(\delta\beta(\theta_1,\phi_1)+\delta\gamma(\theta_1,\phi_1)\right)\nonumber\\
&&+f(\alpha,\beta(\theta_1,\phi_1),\gamma(\theta_1,\phi_1))\nonumber\\
&&=1+2\Delta^{(0i)}\delta\Omega^i(\theta_1,\phi_1)-\Delta^{ij}\delta(\Omega^i\Omega^j)+f\big|_{\theta_1,\phi_1},~~~
\end{eqnarray}
where
$\delta\Omega^i(\theta_1,\phi_1)\equiv\Omega^i(\theta_1,\phi_1)-\Omega^i(\theta_0,\phi_0)$.
$2\delta\Omega^i(\theta_1,\phi_1)$ and $\delta(\Omega^i\Omega^j)$
are $O(1)$, so $\Delta^{(0i)}\sim 10^{-5}$ and $\Delta^{ij}\sim
10^{-5}$ to give the OPERA speed anomaly of $\sim10^{-5}$. If
$(\theta_1,\phi_1)=(\theta_0,\phi_0)$, then
$f\big|_{\theta_1,\phi_1}\sim 10^{-5}$. The illustration is shown in
Fig.~\ref{opera_illustration}.

It has been reported~\cite{ICARUS} by the ICARUS Collaboration that
there is no evidence for the analogues Cherenkov radiation of muon
neutrinos from CERN to the LNGS, where the OPERA experiment is also
performed. If taking the arguments of Cohen-Glashow and Bi {\it et
al.} as true, then one must refute the OPERA result as
Ref.~\cite{ICARUS} did. However, we take the ICARUS result as a
support of our argument on the forbidding of these Cherenkov-like
processes, rather than a refutation of the OPERA result. Therefore
our argument can accommodate both the OPERA and the ICARUS
experiments, whereas one must refute the OPERA result or the ICARUS
result based on the arguments for these Cherenkov-like processes.

In summary, we get that even if the Lorentz violation parameters
$\Delta^{0i}_{\nu_\mu}$, $\Delta^{i0}_{\nu_\mu}$ and
$\Delta^{ij}_{\nu_\mu}$ are $\sim10^{-5}$ for the OPERA experiment,
muon neutrinos are still able to be produced through processes
$\pi/K\rightarrow\mu +\nu_\mu$. After being generated, muon
neutrinos are still able to be free from the Cherenkov analogous
processes $\nu_\mu\rightarrow\nu_\mu+e^{+}+e^{-}$ and etc,
propagating from CERN to Gran Sasso, if the Lorentz violation is
covariant with the muon neutrino when the Cherenkov-like processes
happen. So the argument of Ref.~\cite{Glashow11} is just among one
of the theoretically assumption-dependent possibilities. The
``obvious" forbidding to the OPERA experiment of muon neutrinos is
not so obvious. Our argument not only manifest the theoretical
rationality for the Lorentz violation of muon neutrinos in the OPERA
experiment, but can also accommodate both the OPERA and the ICARUS
results. On the other hand, we agree that the superluminality of
muon neutrinos in the OPERA experiment still needs to be checked by
further experiments.

\begin{acknowledgements}
The work is supported by National Natural Science Foundation of
China (Nos. 10975003, 11021092, 11035003, and 11120101004), by
the Key Grant Project of Chinese Ministry of Education (No. 305001),
and by the Research Fund for the Doctoral Program of Higher
Education, China.
\end{acknowledgements}

\end{document}